\documentclass{ifacconf}

\usepackage[british]{babel}
\usepackage{amsmath,amssymb,amsfonts}
\usepackage{graphicx}
\usepackage{textcomp}
\usepackage{float}
\usepackage{xcolor}

\usepackage{mathrsfs}
\usepackage{lipsum}
\usepackage{natbib}
\usepackage{algorithm}

\usepackage{algorithm}
\usepackage{algpseudocode}

\newif\ifshowremove
\showremovetrue

\newcommand{\Borel}{\mathcal{B}}
\renewcommand{\Pr}{\mathbb{P}}

\newcommand{\X}{\mathbb{X}}

\newcommand{\true}{\top}

\makeatletter
\DeclareRobustCommand{\qed}{%
	\ifmmode 
	\else \leavevmode\unskip\penalty9999 \hbox{}\nobreak\hfill
	\fi
	\quad\hbox{\qedsymbol}}
\newcommand{\openbox}{\leavevmode
	\hbox to.77778em{%
		\hfil\vrule
		\vbox to.675em{\hrule width.6em\vfil\hrule}%
		\vrule\hfil}}
\newcommand{\qedsymbol}{\openbox}
\newenvironment{proof}[1][\proofname]{\par
	\normalfont
	\topsep6\p@\@plus6\p@ \trivlist
	\item[\hskip\labelsep\itshape
	#1.]\ignorespaces
}{%
	\qed\endtrivlist
}
\newcommand{\proofname}{Proof}
\makeatother

\newtheorem{theorem}{Theorem}

\newtheorem{definition}[theorem]{\noindent Definition}
\newtheorem{remark}{Remark}
\newtheorem{lemma}[theorem]{Lemma}
\newtheorem{assumption}{Assumption}

\def\mycmd{0} 

\allowdisplaybreaks

\begin{document}
	\begin{frontmatter}

		\title{Risk-Aware MPC for Stochastic Systems with Runtime Temporal Logics}

		\thanks[footnoteinfo]{This work is supported by the Dutch NWO Veni project CODEC under grant number 18244 and the European project SymAware under the grant number 101070802.}

		\author[First]{M.H.W. Engelaar}
		\author[First]{Z. Zhang}
		\author[First]{M. Lazar}
		\author[First]{S. Haesaert}

		\address[First]{Department of Electrical Engineering (Control Systems Group), Eindhoven University of Technology, The Netherlands. Emails:\{m.h.w.engelaar, z.zhang3, m.lazar, s.haesaert\}@tue.nl}

		\begin{abstract}                
			: This paper concerns the risk-aware control of stochastic systems with temporal logic specifications dynamically assigned during runtime. Conventional risk-aware control typically assumes that all specifications are predefined and remain unchanged during runtime. In this paper, we propose a novel, provably correct model predictive control scheme for linear systems with additive unbounded stochastic disturbances that dynamically evaluates the feasibility of runtime signal temporal logic specifications and automatically reschedules the control inputs accordingly. The control method guarantees the probabilistic satisfaction of newly accepted specifications without sacrificing the satisfaction of the previously accepted ones. The proposed control method is validated by a robotic motion planning case study. 
			\vspace{-1.0em}
		\end{abstract}
	
			\vspace{-0.2em}
	

	\end{frontmatter}

	\section{introduction}
	Practical engineering systems are required to accomplish certain desired tasks with restricted risk levels. This can be achieved by risk-aware control incorporating stochastic uncertainties, which renders a stochastic planning problem with probabilistic risk constraints~\citep{sadigh2016safe}. For stochastic systems with signal temporal logic (STL) specifications, control synthesis has been performed in~\citet{Fara2018} by underapproximating the fixed chance constraints using linear inequalities. Similarly, for dynamic systems with bounded disturbances, control synthesis has been performed based on the worst-case optimization~\citep{Fara2015, Rama2015,Sadr2019}. For probability STL (PrSTL), one can convert the probabilistic satisfaction of specifications to a Boolean combination of risk constraints~\citep{sadigh2016safe}. For example, this method has been used in~\citet{Mehr2017} to solve the ramp merging problem. 

	\vspace{-0.5em}

	\begin{figure}[htp]
		\includegraphics[width=0.8\columnwidth]{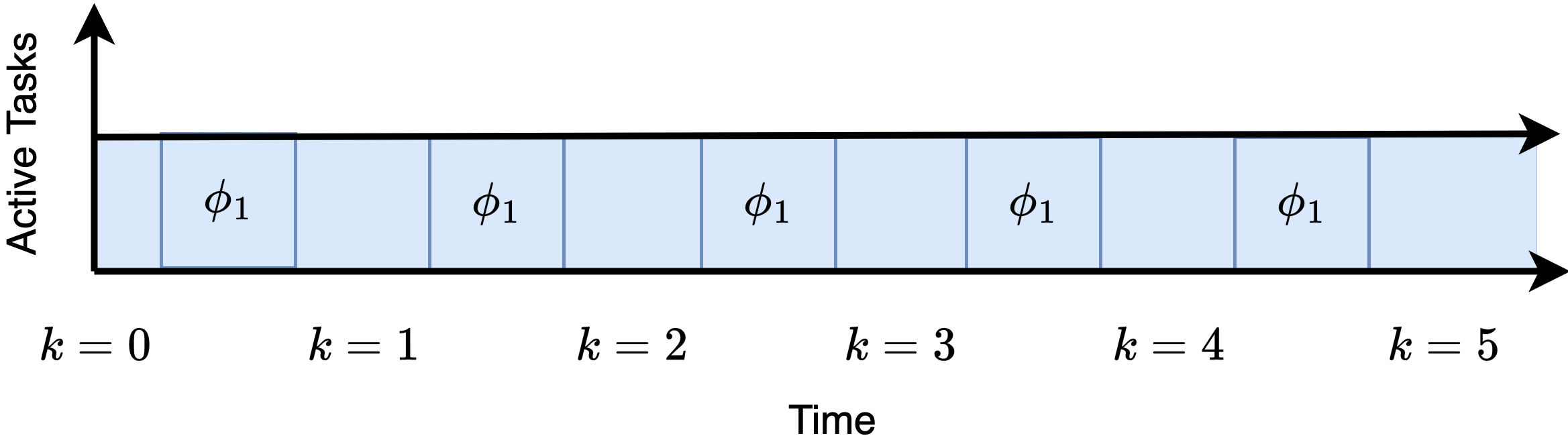}
		\centering
		\vspace{-1em}
		\caption{ Illustration of the classical specification problem.}
		\label{Fig:ProblemIllustration1}
	\end{figure}

	\vspace{-0.5em}

\begin{figure}[htp]
		\includegraphics[width=0.78\columnwidth]{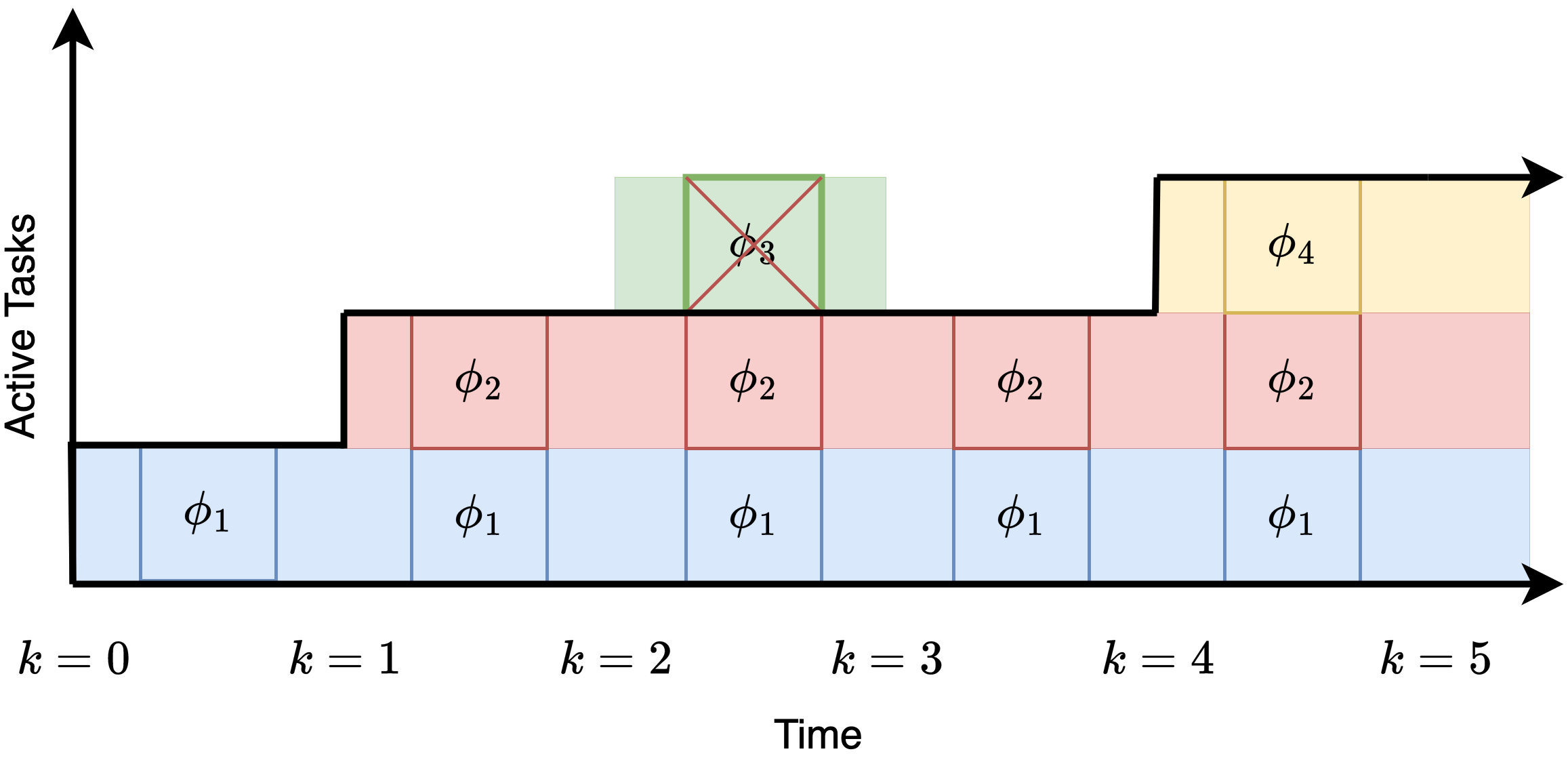}
		\centering
		\vspace{-1em}
		\caption{ Illustration of the dynamic specification problem. Here, a new task is accepted at times $k=0$, $k=1$, and $k=4$, while at time $k=2$, a task is rejected.}
		\label{Fig:ProblemIllustration2}
	\end{figure}

	Common amongst all previously mentioned works is the assumption of predefined specifications (or tasks), as illustrated in Fig. \ref{Fig:ProblemIllustration1}. Moreover, predefining specifications extends to most work on conventional risk-aware control of stochastic systems. Nevertheless, practical applications in multi-robot systems and human-robot collaboration often need to dynamically allocate tasks during runtime~\citep{zhu2006neural, bruno2018dynamic}. For example, a restaurant service robot delivering a dish may receive another delivery task at any time. It then needs to evaluate whether taking the order affects the accomplishment of the current delivery. As shown in Fig. \ref{Fig:ProblemIllustration2}, for a general control system, this indicates that it should automatically reschedule its current control policy by evaluating the risks of newly assigned specifications during runtime. To our knowledge, such control schemes have yet to be introduced in the literature, although some inspiring work exists. A shrinking-horizon model predictive control (MPC) method has been provided in~\citet{Fara2018} to solve closed-loop risk-aware control of stochastic systems with temporal logic specifications. In~\citet{mustafa2023probabilistic}, multiple risk bounds are evaluated from a set of control policies where the best policy with the least conservativeness is selected. The work in~\citet{Fara2015} proposes a scheme to accept or reject a specification by evaluating its feasibility. In \citet{Hewi2018}, tube-based stochastic MPC has been used to synthesize controllers for predefined probabilistic set constraints. In \citet{Enge2023_2}, probabilistic reachable tubes have been used to optimize risk bounds over predefined probabilistic set constraints. These approaches and concepts provide inspiring ideas for the risk re-evaluation and control rescheduling of stochastic systems with runtime specifications.

	Motivated by the above open problem, this paper develops a risk-aware rescheduling control scheme for stochastic linear systems with runtime STL specifications. Inspired by the shrinking horizon MPC approach of~\citet{Fara2018} and probabilistic reachable sets of \citet{Hewi2018}, both of which assume predefined constraints, we develop a tube-based MPC scheme that allows for dynamically assigned STL specifications. We use probabilistic reachable tubes to re-calculate the risk bounds of the probabilistic constraints and reschedule the control input to satisfy the newly assigned specifications while retaining the satisfaction of the previously accepted specifications, with the liberty of rejection. Our contributions include control synthesis for dynamically assigned STL specifications, the recursive feasibility of the control scheme and proof of open-loop constraint satisfaction. A two-dimensional vehicle motion planning case with dynamically assigned targets is used to validate the efficacy of our method.
	\if\mycmd0
	Proofs and detailed derivations can be found in the Appendix. 
	\else
	Proofs and detailed derivations can be found in the extended report \citep{engelaar2024risk}.
	\fi



	\section{Preliminaries and Problem Statement} \label{Sec:ProSet}

%
%

	For a given probability measure $\mathbb P$ defined over Borel measurable space $(\X,\mathcal{B}(\X))$, we denote the probability of an event $\mathcal{A} \in \Borel(\X)$ as $\mathbb{P}(\mathcal{A})$. In this paper, we will work with Euclidean spaces and Borel measurability. Further details on measurability are omitted, and we refer the interested reader to \citet{Bert1996}. Additionally, any half-space $H \subset \mathbb{R}^n$ is defined as $H:=\{x \in \mathbb{R}^n \mid g^Tx\geq b\}$ with $g \in \mathbb{R}^n$ and $b \in \mathbb{R}$. A polyhedron is the intersection of finitely many half-spaces, also denoted as $H:=\{x \in \mathbb{R}^n \mid Gx\geq b\}$ with $G \in \mathbb{R}^{q\times n}$ and $b \in \mathbb{R}^q$. The 2-norm is defined by $||x||=\sqrt{x^Tx}$ with $x \in \mathbb{R}^n$. The identity matrix is denoted by $I_n \in \mathbb{R}^{n \times n}$. The vectors of all elements one and zero are denoted by $\boldsymbol{1}_{n} \in \mathbb{R}^{n}$ and $\boldsymbol 0_n \in \mathbb{R}^n$, respectively. If the context is clear, they will be denoted by $\boldsymbol 1$ and $\boldsymbol 0$.


	\subsection{Class of Systems}

	In this paper, we consider systems with linear time-invariant dynamics with additive noise, given by
	\begin{equation} \label{Sys}
		x(k+1)=Ax(k)+Bu(k)+w(k),
	\end{equation}
	where matrix pair $(A,B)$ is stabilizable, $x \in \mathbb{X} \subseteq \mathbb{R}^n$ is the state of the system, $x(0) \in \mathbb{X}$ is an initial state, $u \in \mathbb{U} \subseteq \mathbb{R}^m$ is the input of the system and $w \in \mathbb{R}^n$ is an independent, identically distributed noise disturbance with distribution $\mathcal{Q}_w$, i.e.,  $w(k) \sim \mathcal{Q}_w$, which can have infinite support. We will assume that the distribution has at least a known mean and variance, with the latter assumed to be strictly positive definite. Additionally, we assume that the distribution is central convex unimodal\footnote{$\mathcal{Q}_w$ is in the closed convex hull of all uniform distributions on symmetric compact convex bodies in $\mathbb{R}^n$ (c.f. \citet[Def. 3.1]{Dhar1976}).}.

	To control the system, we define a sequence of policies $$\boldsymbol f:= \{f_0, f_1,\dots\},$$ such that $f_k: \mathbb{H}_k \to \mathbb{U}$ maps the history of states and inputs to inputs. Here $\mathbb{H}_k:=(\mathbb{X} \times \mathbb{U})^k \times \mathbb{X}$ with element $\eta(k):=(x(0),u(0), \cdots, u(k-1), x(k))$. By implementing controller $\boldsymbol f$ upon system \eqref{Sys}, we obtain its controlled form for which the control input satisfies the feedback law $u(k)= f_k(\eta(k))$ with $\eta(k) \in \mathbb{H}_k$. We indicate the input sequence of system \eqref{Sys} by $\boldsymbol u:=\{u(0),u(1),\dots\}$ and define its executions as sequences of states $\boldsymbol x:=\{x(0),x(1),\dots\}$, referred to as signals. Any signal $\boldsymbol{x}$ can be interpreted as a realization of the probability distribution induced by implementing controller $\boldsymbol{f}$, denoted by $\boldsymbol{x} \sim \mathbb{P}_{\boldsymbol{f}}$. Similarly, the suffix fragment $\vec{\boldsymbol{x}}_k=\{x(k), x(k+1), \dots\}$ of a signal $\boldsymbol{x}$ can be interpreted as a realization $\boldsymbol{\vec x}_k \sim \mathbb{P}_{\boldsymbol{f},k}$. We define the segment and signal element of any signal $\boldsymbol x$ by $\boldsymbol x_{[i,j]}:=\{x(i), \ldots, x(j)\}$ with $i\leq j$ and $\boldsymbol x(k):=\boldsymbol x_{[k,k]}$.


	\subsection{Signal Temporal Logic}

	To mathematically describe tasks, e.g., reaching a target within a given time frame, we consider specifications given by signal temporal logic. Here, we assume that all STL specifications adhere to the \emph{negation normal form} (NNF), see \citet{Rama2014}. This assumption does not restrict the overall framework since \citet{Sadr2015} proves that every STL specification can be rewritten into the negation normal form.

	STL consists of predicates $\mu$ that are either true ($\top$) or false ($\bot$). Each predicate $\mu$ is described by a function $h:\mathbb{R}^n \to \mathbb{R}^q$ as follows
	\begin{equation}
		\mu:=\begin{cases}
			\top \text{ if } \ h(x) \geq \boldsymbol 0 \\
			\bot \text{ else}.
		\end{cases}
	\end{equation}
	We will assume that all predicate functions are linear affine functions of the form $h(x)=Gx - b$ with $G \in \mathbb{R}^{q \times n}$, $b\in \mathbb{R}^q$ and $||g_j||=1$ for all $j \in \{1, \cdots, q\}$, where $g_j$ is the $j$-th row of $G$. Notice that satisfaction of the predicate $\mu$ is defined by whether $x$ is contained within the polyhedron $H_{\mu}:=\{ x \in \mathbb{R}^n \mid G x \geq b\}$. Since all rows of $G$ are normalized, we denote the polyhedron $H_{\mu}$ as normalized. Let the set of all predicates be given by $\mathcal{P}$. The syntax of STL will be given by
	$$\phi::= \true \mid \mu \mid \lnot \mu \mid \phi_1 \wedge \phi_2 \mid \phi_1 \vee \phi_2 \mid \square_{[a,b]} \phi \mid \phi_1 \ U_{[a,b]} \ \phi_2, $$
	where $\mu \in \mathcal{P}$, $\phi$, $\phi_1$ and $\phi_2$ are STL formula, $a,b \in \mathbb{N}$, and $a \leq b$. The semantics are given next, where $\boldsymbol{\vec x}_k \vDash \phi$ denotes the suffix of signal $\boldsymbol x$ satisfying specification $\phi$.
	\begin{definition}
		The STL semantics are recursively given by:
		\begin{align*}
			&	 \boldsymbol{x}_k \vDash \mu & \iff & h({\boldsymbol x(k)})\geq \boldsymbol 0  \\
			&	 \boldsymbol{x}_k \vDash \lnot \mu & \iff & h({\boldsymbol x(k)}) \ngeq \boldsymbol 0  \\
			&	 \boldsymbol{\vec x}_k \vDash \phi_1 \wedge \phi_2 & \iff & \boldsymbol{\vec x}_k \vDash \phi_1 \text{ and } \boldsymbol x_k \vDash \phi_2 \\
			&	 \boldsymbol{\vec x}_k \vDash \phi_1 \vee \phi_2 & \iff & \boldsymbol{\vec x}_k \vDash \phi_1 \text{ or } \boldsymbol x_k \vDash \phi_2 \\
			&	 \boldsymbol{\vec x}_k \vDash \square_{[a,b]} \ \phi & \iff & \forall k' \in [k+a,k+b]: \boldsymbol{\vec x}_{k'} \vDash \phi \\
			&	 \boldsymbol{\vec x}_k \vDash \phi_1 \ U_{[a,b]} \ \phi_2 & \iff & \exists k_1 \in [k+a,k+b]: \boldsymbol{\vec x}_{k_1} \vDash \phi_2 \\
			&	&  &\hspace{-.75cm}  \text{ and }  \forall k_2 \in [k,k_1], \ \boldsymbol{\vec x}_{k_2} \vDash \phi_1.
		\end{align*}
	\end{definition}
	Here, we introduced a slight abuse of the notation as $[a,b]$ is used to describe the set of all natural numbers within the real interval $[a,b]$. Additional operators can be derived such as the eventually-operator $\lozenge_{[a,b]} \phi := \true U_{[a,b]} \phi$.

	Since the dynamical system behaves stochastically, each STL specification can only be satisfied probabilistically. Should a specification be assigned at time $k$, the probability can be determined based on the state measurement $x(k)$, the controller $\boldsymbol f$, and the system dynamics \eqref{Sys}. Accordingly, we consider the probability that suffix fragment $\vec{\boldsymbol x}_k$ satisfies specification $\phi$, given state $x(k)$, denoted by
	\begin{equation}
		\mathbb{P}_{\boldsymbol f}(\phi,k):=\mathbb{P}_{\boldsymbol f, k}(\vec{\boldsymbol x}_k \vDash \phi \mid x(k)).
	\end{equation}
	For each specification $\phi$, we assume that the maximal allowable risk $r_{\phi, \max}$ of not satisfying $\phi$ is given. In the remainder, we will refer to this as the maximal risk.
	


	\subsection{Problem Statement}
	
	We consider a dynamical system \eqref{Sys} that needs to satisfy a sequence of STL specifications. Specifications will be provided in real-time and are assumed to be unknown until the time of assignment. The paper's objective is two-fold. Firstly, to design a control synthesis technique that gives probabilistic guarantees, i.e., satisfaction of any STL specification with respect to its maximal risk. Secondly, to update the control strategy in real-time whenever a new specification is assigned to the dynamical system. Here, a new specification may be rejected if necessary.

	We formalize the above problem description into the following problem statement. Here, we consider the dynamical system \eqref{Sys} with noise distribution $\mathcal{Q}_w$ and state measurement $x(k)$, at each time step $k \in \mathbb{N}$.




	\textbf{Problem Statement.}
	Given a sequence of STL specifications $\boldsymbol{\phi}:=\{\phi_1, \cdots, \phi_t\}$ assigned, respectively, at time instances $\{k_1, \cdots, k_t\}$ with $k_i<k_j$ for $i<j$, with maximal risk $r_{\phi_i,\max}$, develop a method for \emph{updating}, at each time step $k \in \mathbb{N}$, suffix $\boldsymbol f_k$ such that $\mathbb{P}_{\boldsymbol f}(\phi_i, k_i)\geq1-r_{\phi_i,\max}$ if $k_i \leq k$, where $\phi_i$ with $k_i=k$ is allowed to be rejected.

\begin{figure}[htp]
		\centering
		\includegraphics[width=0.8\columnwidth]{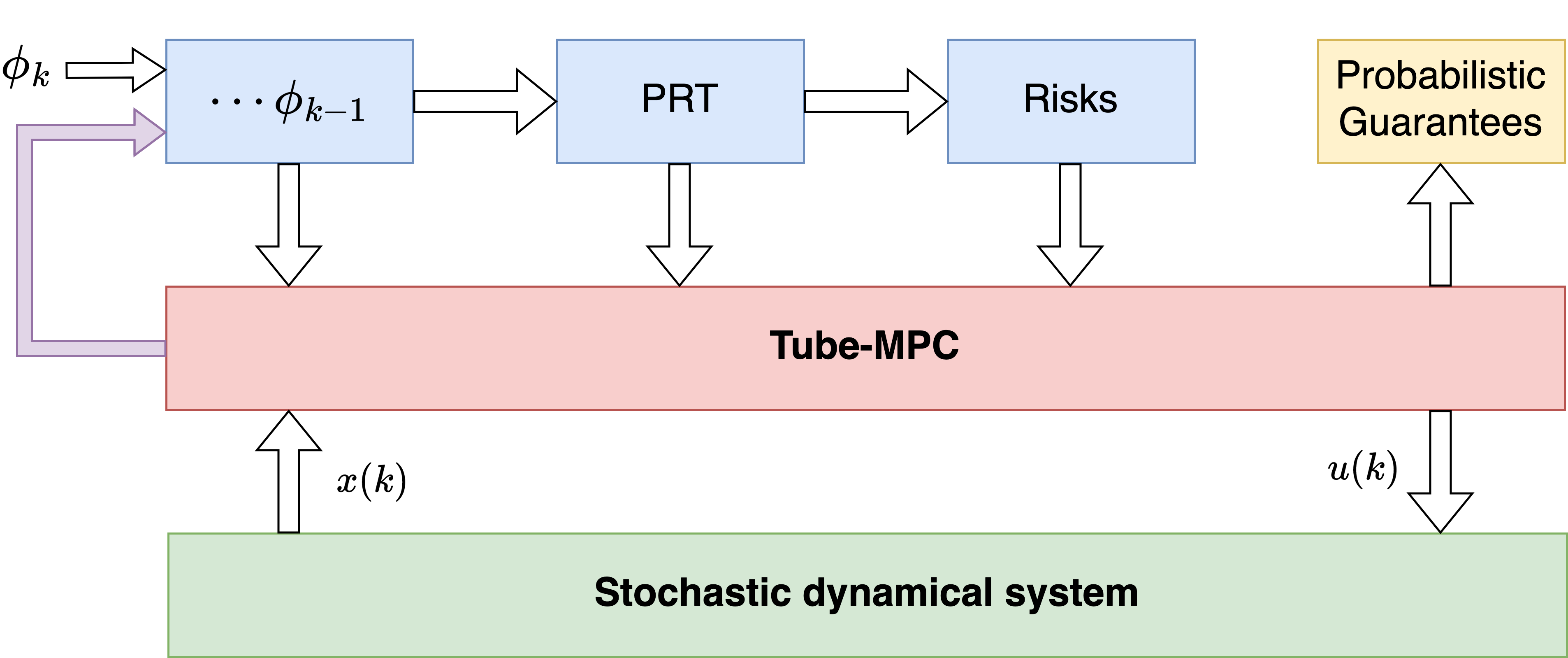}
		\vspace{-0.6em}
		\caption{ The developed approach: real-time specifications, PRTs and risks are dynamically updated and integrated within a deterministic tube-MPC problem, which yields a control update.}
		\label{Fig:Approach}
	\end{figure}

	\textbf{Approach.}
	To solve the problem statement, we consider the approach in Fig. \ref{Fig:Approach}. First, we reformulate STL specifications into mixed integer constraints. Next, a probabilistic reachable tube (PRT) is fitted inside these constraints. Here, the tube can upper bound the risk that states are not contained within the tube. By constraining the risks, a tube-based MPC problem is defined at each time instance, which reschedules the controller, decides whether a new specification can be accepted and updates the guarantees.

	\section{Risk-Aware Tube-MPC design with Rescheduling} \label{Sec:Main}

	In this section, we first define probabilistic reachable tubes and utilize those to relate each STL specification to its maximal risk via mixed integer (non-)linear constraints. Second, we utilize the previously mentioned relation to formulate a tube-based MPC problem, which will reschedule the controller per the requirements of the problem statement. Third, we prove recursive feasibility of the overall scheme and prove open-loop constraint satisfaction.

	In this paper, we decompose dynamics \eqref{Sys} into a nominal and an error part \citep{Hewi2018}. The nominal dynamics, denoted as $z$,  contain no stochasticity, and the (stochastic) error dynamics, denoted as $e$, are autonomous. This yields
	\begin{subequations} \label{Eq:NomErr}
		\begin{align}
			z(k+1) & = Az(k)+Bv(k), \label{Eq:Nom}\\
			e(k+1) &= A_Ke(k)+w(k), \label{ErrorPart}\\
			\mbox{with  }
			x(k) &= z(k)+ e(k),\\
			u(k) &= v(k) +Ke(k).
		\end{align}
	\end{subequations}
	Here $A_K=A+BK$ and $K$ is a stabilizing feedback gain meant to keep the error $e$ small.


	\subsection{Probabilistic Reachable Tubes}

	We define probabilistic reachable sets as follows, similar to \citet{ Hewi2018}.
	\begin{definition}[Probabilistic Reachable Sets] \label{Def:PRS}
		A probabilis-tic reachable set (PRS) of dynamic probability level $p$ for error dynamics \eqref{ErrorPart}, denoted by $\mathcal{R}^p$, is a set that satisfies
		\begin{equation}\label{Eq:PRT}
			e(0) = \boldsymbol 0 \Rightarrow \Pr(e(k) \in \mathcal{R}^p) \geq p, \ \forall k \geq 0.
		\end{equation}
	\end{definition}
	Definition \ref{Def:PRS} states that a PRS of probability level $p$ will contain at any time instance $k$ the accumulating error $e(k)$ with at least probability $p$, if the initial error satisfies $e(0)=\boldsymbol 0$. 
	Multiple representations for probabilistic reachable sets exist, including the ellipsoidal representation. For simplicity, we will make the following assumption.

	\begin{assumption} \label{Ass:Zero}
		Distribution $\mathcal{Q}_w$ has zero mean.
	\end{assumption}

	The above assumption is not necessary to obtain an ellipsoidal representation but will simplify most of the computations. The ellipsoidal representation is obtained from the multivariable Chebyshev inequality; details are given in \citet{Hewi2018,farina2016stochastic}. Under Ass. \ref{Ass:Zero}, the ellipsoidal representation is given by
	\begin{equation}\label{EllipNot}
		\mathcal{R}^p= \{x \in \mathbb{R}^{n} \mid x^T \Sigma_{\infty}^{-1}x \leq \tilde p\},
	\end{equation}
	where $\Sigma_{\infty}$ solves the Lyapunov equation
	\begin{equation} \label{Eq:Lyap}
		A_K\Sigma_{\infty}A_K^T+\text{var}(\mathcal{Q}_w)=\Sigma_{\infty},
	\end{equation}
	$p$ is the probability level, and
	\begin{equation} \label{Eq:RelRadProb}
		\textstyle \tilde p=\frac{n}{1-p}.
	\end{equation}
	Since the variance of $\mathcal{Q}_w$ is assumed to be strictly positive definite and $A_K$ is stable, Lyapunov theory states that $\Sigma_{\infty}$ is strictly positive definite. Hence, in the sequel, we consider the transformed state and constraints for which $\Sigma_{\infty}=I_n$. These can be obtained by a simple coordinate transformation $y=\Sigma_{\infty}^{-\frac{1}{2}}x$. We call the transformed dynamics \textit{the normalized dynamics}, and in the remainder, we maintain our notations as if the dynamics are normalized. Notice that each PRS is now a hypersphere with radius $\sqrt{\tilde p}$. Accordingly, Eq. \eqref{Eq:RelRadProb} is now a relation between the radius and the probability level of each PRS.

	\begin{remark}
		In the case of Gaussian disturbance, a less conservative relation exists, compared to Eq. \eqref{Eq:RelRadProb}.	This alternative relation is given by $\tilde p=\chi^2_{n}(p)$, the inverse cumulative distribution function of the chi-squared distribution with $n$ degrees of freedom.
	\end{remark}

	By defining a PRS $\mathcal{R}^{p_k}$ at each time instance $k \in \{0, \cdots, N\}$, we obtain a finite sequence of PRS called a probabilistic reachable tube $\mathcal{R}:=\{\mathcal{R}^{p_0}, \cdots, \mathcal{R}^{p_N}\}$. A probabilistic reachable tube can have either constant or varying diameter, with the former only if $p(i)=p(j)$ for all $i\leq j$. We define the risk level $r_k$ as an upper bound on the probability that $\mathcal{R}^{p_k}$ does not contain the error $e(k)$.


	\subsection{Risk-Aware Tube-MPC Constraints Design}
	
	This subsection aims to relate each STL specification $\phi$ to its maximal risk $r_{\phi, \max}$ via mixed integer (non-) linear constraints. To achieve this, we relate each STL specification $\phi$ assigned at time $k$, by means of a PRT, to the risk $r_\phi^k$ of failing specification $\phi$ assigned at time $k$. Subsequently, the desired relation can be derived by imposing that $r_{\phi}^k \leq r_{\phi,\max}$.
	
	We first define the active horizon $\mathcal{H}_{\phi}^k$  of any specification $\phi$ as the set of all time instances needed to evaluate $\boldsymbol x_k \vDash \phi$.
	\begin{definition}
		The active horizon is recursively given by:
		\begin{align*}
			&	 \boldsymbol{\vec x}_k \vDash \true & \iff & \mathcal{H}_{\true}^k = \emptyset  \\
			&	 \boldsymbol{\vec x}_k \vDash \mu & \implies & \mathcal{H}_{\mu}^k = \{k\}  \\
			&	 \boldsymbol{\vec x}_k \vDash \ \lnot \mu & \implies & \mathcal{H}_{\lnot \mu}^k = \{k\}  \\
			&	 \boldsymbol{\vec x}_k \vDash \phi_1 \wedge \phi_2 & \implies & \mathcal{H}_{\phi_1 \wedge \phi_2}^k=\mathcal{H}_{\phi_1}^k \cup \mathcal{H}_{\phi_2}^k  \\
			&	 \boldsymbol{\vec x}_k \vDash \phi_1 \vee \phi_2 & \implies & \mathcal{H}_{\phi_1 \vee \phi_2}^k=\mathcal{H}_{\phi_1}^k \cup \mathcal{H}_{\phi_2}^k  \\
			&	 \boldsymbol{\vec x}_k \vDash \square_{[a,b]} \ \phi & \implies & \mathcal{H}_{\square_{[a,b]} \phi}^k=\cup_{i=k+a}^{k+b}\mathcal{H}_{\phi}^i\\
			&	 \boldsymbol{\vec x}_k \vDash \phi_1 \ U_{[a,b]} \ \phi_2 & \implies & \mathcal{H}_{\phi_1 \ U_{[a,b]} \ \phi_2}= \mathcal{H}_1 \cup \mathcal{H}_2\\
			&	&  &\hspace{-3cm}  \text{ with } \mathcal{H}_1=\cup_{i=k}^{k+b}\mathcal{H}_{\phi_1}^i \text{ and } \mathcal{H}_2=\cup_{j=k+a}^{k+b}\mathcal{H}_{\phi_2}^j.
		\end{align*}
	\end{definition}
	\vspace{-0.7em}
	It should be noted that any active horizon is a finite subset of $\mathbb{N}$. Notice that the suffix $\boldsymbol x_{N+1}$ with $N=\max{\mathcal{H}_\phi^k}$, does not influence the  satisfaction of $\boldsymbol x_k \vDash \phi$. Hence, in the remainder of the paper, we consider only the segment $\boldsymbol x_{[k,N]}:=\{x(k), \cdots, x(N)\}$ and note that $\mathbb{P}_{\boldsymbol f}(\phi,k)=\mathbb{P}_{\boldsymbol f, k}(\vec{\boldsymbol x}_k \vDash \phi \mid x(k))=\mathbb{P}_{\boldsymbol f, k}(\vec{\boldsymbol x}_{[k,N]} \vDash \phi \mid x(k))$.

	\begin{remark}	
		The value of $N$ defined above is closely related to the length $L(\cdot)$ of an STL specification, formally defined in~\citet{maler2004monitoring}. Specifically, for a specification $\phi$ assigned at time $k$, we have $L(\phi) = N-k$.
	\end{remark}

	Given an STL specification $\phi$ assigned at time $k$, \citet{Rama2014} explains that mixed integer linear constraints can represent any STL specification in negation normal form. As can be observed in the following equations, for each time instance $i\in \mathcal{H}_{\phi}^k$, these STL specifications define a linear constraint that is parametrized with binary variable $s$, as follows
	\begin{subequations}\label{Eq:STLCon}
		\begin{align}
			G(i)x(i)+MF(i)s &\geq b(M,i) + \rho(i)\boldsymbol{1},\\
			Cs \geq d, \ \epsilon \leq \rho(i) &\leq \frac{M}{2}, \ \forall i \in \mathcal{H}_{\phi}^k, \label{Eq:STLCon_b}
		\end{align}
	\end{subequations}
	where $G(i)$, $E(i)$, $C$ and $b(M,i)$, $d$ are, respectively, matrices and vectors of appropriate sizes; $M$ is a large positive number; $\epsilon$ is a small positive number; and $\rho(i)>0 \in \mathbb{R}$ is a positive real number. 
	\if\mycmd0
	For more information on the derivation, consider the Appendix.
	\else
	For more information on the derivation, consider the appendix of the extended version \citep{engelaar2024risk}.
	\fi
	
	The following theorem quantifies the risk $r_{\phi}^k$ via constraints \eqref{Eq:STLCon}. 
	\if\mycmd0
	Both an explanation and proof can be found in the Appendix.
	\else
	Both an explanation and proof can be found in the appendix of the extended version \citep{engelaar2024risk}.
	\fi
	
	\begin{theorem} \label{Thm:Risk}
		Let specification $\phi$ be assigned at time $k$ to the system 	\eqref{Eq:NomErr} with normalized dynamics. If the nominal trajectory $\{z(k), \cdots, z(N)\}$ with $z(k)=x(k)$ and nominal controls $\{v(k), \cdots, v(N-1)\}$ satisfies \eqref{Eq:STLCon} with $\{\rho(k), \cdots, \rho(N)\}$ strictly positive, than $\mathbb{P}_{\boldsymbol f}(\phi,k)=\mathbb{P}_{\boldsymbol f, k}(\vec{\boldsymbol x}_k \vDash \phi \mid x(k))\geq 1-r_{\phi}^k,$
		where risk $r_{\phi}^k$ is at most
		\begin{equation}\label{Eq:Risk}
			\textstyle r_{\phi}^k =\sum_{i\in \mathcal{H}_\phi^k \backslash \{k\} } \frac{n}{\rho(i)^2}.
		\end{equation}
	\end{theorem}


	\subsection{Risk-Aware Tube-MPC Rescheduling}
	
	In this subsection, we will design a tube-based MPC scheme to update the controller during runtime. We consider two distinct scenarios, with the latter building upon the former. These two scenarios are the absence and the presence of a new specification.
		
	For each specification $\psi$, we denote with $k_{\psi}$ the time at which the specification was accepted. We define the set of accepted specifications at time $k$ by $\mathcal{P}^k:=\{\psi \mid k_{\psi} \leq k\}$.
	
	\vspace{-0.33em}
	
	\textbf{Absence of new Specification. }
	Without loss of generality, we assume a controller exists at time $k-1$. In the absence of a new specification at time $k$, the set $\mathcal{P}^k$ contains only previously accepted specifications. The objective is to update the controller while ensuring satisfaction for all $\psi \in \mathcal{P}^k$. Utilizing the relation between specifications and their maximal risks given by \eqref{Eq:STLCon}, \eqref{Eq:Risk} and $r_{\psi}^{k_{\psi}} \leq r_{\psi,\max}$, we can guarantee satisfaction, if the nominal dynamics satisfy the aforementioned constraints, following Thm. \ref{Thm:Risk}.
	
	We develop a tube-based MPC algorithm that at each time instance $k$ recomputes the optimal nominal trajectory $\boldsymbol{z}_{[k, N]}$ and the optimal nominal input $\boldsymbol{v}_{[k,N-1]}$ ensuring satisfaction of the previously mentioned constraints. We assume given $x(k)$ together with $\boldsymbol z_{[0,k]}$, $\boldsymbol v_{[0,k-1]}$, $\boldsymbol \rho_{[0,k]}$ and $\boldsymbol r_{[0,k]}$ computed at time $k-1$. For all $\psi\in \mathcal{P}^k$, we assume $\max\mathcal{H}_{\psi}^{k_{\psi}} \leq N$. This assumption will simplify most of the computations; however, all results remain valid should the horizon be allowed to change during runtime.
	
	We consider the following tube-based MPC problem
	\begin{subequations}\label{Eq:TMPC}
		\begin{align}
			\min_{\Xi}  \ &   J(\boldsymbol z_{[0,N]}, \boldsymbol v_{[0,N-1]}, \boldsymbol r_{[0,N]})+Ma(k),\\
			\text{s.t. }	\ &z(k) = (1-a(k))x(k)+a(k)z_{k}, \ a(k) \in \{0, 1\},\!\! \label{Eq:TMPC_meas}\\
			& z(i+1)= Az(i)+Bv(i), \forall i \in \{k, \cdots, N-1\},\!\! \label{Eq:TMPC_sys} \\[0.4em]
			&G(j)z(j)+MF(j)s(k) \geq b(M,j) + \rho(j)\boldsymbol{1}, \label{Eq:TMPC_SpecOld}\\
			&Cs(k) \geq d, \ \epsilon \leq \rho(j)\leq \textstyle \frac{M}{2},\ \forall  j\in \{0, \cdots, N\}, \label{Eq:TMPC_SpecCon}\\[0.4em]
			&n = \rho(j)^2 r(j), \hspace{2cm}\forall  j\in \{0, \cdots, N\}, \label{Eq:TMPC_ConRiskPRT}\\
			& \textstyle \sum_{j \in \mathcal{H}_\psi^{k_\psi}\setminus \{k_\psi\}} \ r(j) \leq  r_{\psi,\max},   \hspace{1.0cm} \forall \psi \in  \mathcal{P}^k,  \label{Eq:TMPC_ProbGuar}
		\end{align}
	\end{subequations}
	where $\Xi:=\{\boldsymbol{z}_{[k, N]}, \boldsymbol{v}_{[k,N-1]}, \boldsymbol{\rho}_{[k+1,N]}, \boldsymbol{r}_{[k+1,N]}, s(k), a(k)\}$, $z_k$ is the last element of $\boldsymbol z_{[0,k]}$,  $J$  is either a linear or quadratic function, and the constraints \eqref{Eq:TMPC_SpecOld}-\eqref{Eq:TMPC_ProbGuar} are obtained from $\wedge_{\psi \in \mathcal{P}^k}\lozenge_{[k_{\psi}, k_{\psi}]} \psi$,  \eqref{Eq:STLCon}-\eqref{Eq:Risk} and $r_{\psi}^{k_{\psi}} \leq r_{\psi,\max}$.
	
	
	\vspace{-0.33em}
	
	\textbf{Presence of a new Specification. }
	Let $\phi$ be a newly assigned specification at time $k$, $\mathcal P^k = \mathcal P^{k-1}\cup \phi$. Building upon MPC problem \eqref{Eq:TMPC}, $\phi$ introduces additional constraints \eqref{Eq:STLCon}, \eqref{Eq:Risk} and $r_{\phi}^k \leq r_{\phi,\max}$. Here, we allow $\phi$ to be rejected if deemed necessary. Accordingly, we consider the following tube-based MPC problem given by
	\begin{subequations}\label{Eq:TMPCNew}
		\begin{align}
			P_k: \min_{\Xi, c(k)}  \ &  J(\boldsymbol z_{[0,N]}, \boldsymbol v_{[0,N-1]}, \boldsymbol r_{[0,N]})+Ma(k)+Mc(k),\!\! \nonumber\\
			\text{s.t. } & \text{constraints } \eqref{Eq:TMPC_meas}-\eqref{Eq:TMPC_ProbGuar} \text{ for }\psi\in \mathcal P^{k-1}, \nonumber \\[.4em]
			&\bar G(j)z(j)+M\bar F(j)s(k) \geq \bar b(M,j)\label{Eq:TMPC_SpecNew} \\  & \hspace{4cm} + (\rho(j)-c(k)M^2)\boldsymbol{1},\nonumber\\
			&c(k) \geq a(k), \ c(k) \in \{0, 1\} \label{Eq:TMPCNew_Options},\\
			&\textstyle \sum_{j \in \mathcal{H}_\phi^{k}\setminus \{k\}}  r(j) \leq r_{\phi,\max} + c(k)M, \label{Eq:TMPCNew_risk} \!\!\!
		\end{align}
	\end{subequations}
	where the constraint \eqref{Eq:TMPC_SpecNew} is obtained from $\phi$. We remark that $c(k)=1$ implies $\phi$ is rejected, constraint \eqref{Eq:TMPCNew_Options} ensures that a new specification will be accepted only if $z(k)=x(k)$, while inequality \eqref{Eq:TMPCNew_risk} ensures a bound on the risk only if the specification is accepted  ($c(k)=0$). Note that $s(k)$ and \eqref{Eq:TMPC_SpecCon} are obtained from $\wedge_{\psi \in \mathcal{P}^{k-1}\cup \phi}\lozenge_{[k_{\psi}, k_{\psi}]} \psi$.
	
	\vspace{-0.33em}
	
	\textbf{Implementation of MPC Scheme. }		
	At $k=0$, both MPC problems, initialized with $z_0=x(0)$, will find a controller $\boldsymbol f$ composed of the optimal nominal trajectory $\boldsymbol{z}_{[0, N]}$ and the optimal nominal input $\boldsymbol{v}_{[0,N-1]}$. Note that in this case, Eq. \eqref{Eq:TMPC_meas}  is trivially satisfied for $z(0)=z_0=x(0)$. Therefore, Thm. \ref{Thm:Risk} will hold and for each of the specifications $\psi\in \mathcal{P}^0$ the maximal risk $r_{\psi,\max}$ is satisfied if the computed nominal input sequence $\boldsymbol{v}_{[0,N-1]} $ is applied.
	
	At $k>0$, we remark that constraint \eqref{Eq:TMPC_meas} considers either the previously computed nominal state $z(k)$ or the measured state $x(k)$, similar to \citet{Hewi2018}.	In the former case, the previously computed nominal input will be applied and \textit{this will preserve the satisfaction of the maximal risks}. In the latter case, with $z(k)=x(k)$ the MPC algorithms will yield a recomputed optimal nominal trajectory $\boldsymbol{z}_{[k, N]}$ and optimal nominal input $\boldsymbol{v}_{[k,N-1]}$.  Due to Thm. \ref{Thm:Risk}, we can guarantee $\mathbb{P}_{\boldsymbol f}(\psi, k_{\psi})\geq 1- \sum_{j \in \mathcal{H}_\psi^{k_\psi}\setminus \{k_\psi\}} \ r(j)$ for each $\psi\in\mathcal P^k$. This, together with constraint \eqref{Eq:TMPC_ProbGuar}, ensures the maximal risks on previously accepted specifications are maintained.
	
	
	Combining the above MPC problems with computing probabilistic reachable sets, normalizing the dynamics, and reformulating specifications, we get Algorithm \ref{Alg:ConRes} to reschedule the controller and synthesize a control input.
	
	\begin{algorithm}
		
		\caption{Control rescheduling algorithm} \label{Alg:ConRes}
		\begin{algorithmic}[1]
			\State Given: System \eqref{Sys} initialized at $x(0)$
			\State Determine $K$ and compute $\Sigma_{\infty}$ from Eq. \eqref{Eq:Lyap}
			\State Get normalized dynamics of \eqref{Sys}
			\State Set $k=0$, $\mathcal{P}^{-1}=\emptyset$ and $z_0=x(0)$
			
			\For{$k\in\{0,\cdots,N-1\}$}
			
			\State  \textbf{if} no new specification \textbf{then}  
			\underline{go to line \ref{alg:linempc}}\vspace{.2em}
			
			\State  \textit{Rescheduling MPC with old and new specifications:}
			\State ($\phi$, $r_{\phi,\max}$) $\gets$ load new specification
			\State Solve \eqref{Eq:TMPCNew} to get $\boldsymbol z_{[k,N]}$, $\boldsymbol v_{[k,N-1]}$, $c(k)$
			\State	\textbf{if} $c(k)=1$ \textbf{then}  \underline{go to line \ref{alg:linempc2}} \Comment{$\phi$ is rejected}
			\State Set $\mathcal P^k = \mathcal P^{k-1}\cup \phi$, \underline{go  to line \ref{alg:linempc2}} \Comment{$\phi$ is accepted} \vspace{.2em}
			\State  \textit{MPC with only old specifications:}
			
			\State Solve \eqref{Eq:TMPC} to get $\boldsymbol z_{[k,N]}$, $\boldsymbol v_{[k,N-1]}$, \label{alg:linempc}\vspace{.2em}
			\State  \textit{Implement control}
			\State Compute $u(k)$ from Eq. \eqref{Eq:NomErr} \label{alg:linempc2}
			\State Implement $u(k)$, and measure $x(k+1)$
			\EndFor
		\end{algorithmic}
	\end{algorithm}
	
	\subsection{Theoretical Analysis}
	
	As shown next, the tube-MPC \eqref{Eq:TMPCNew} is recursively feasible.
	
	\begin{theorem}
		For all $k \in \{1, \cdots, N\}$, and all $x(k) \in \mathbb{R}^n$, if the tube-MPC problem $P_{k-1}$ defined in \eqref{Eq:TMPCNew} is feasible, then the corresponding problem $P_k$ at the next time instant remains feasible.
	\end{theorem}
	
	\vspace{-0.9em}
	
	\begin{proof}
		Let $\boldsymbol z_{[0,N]}$,  $\boldsymbol v_{[0,N-1]}$, $\boldsymbol \rho_{[0,N]}$ and $\boldsymbol r_{[0,N]}$ be solutions to $P_{k-1}$. Take $a(k)=c(k)=1$ and notice that constraints \eqref{Eq:TMPC_meas}-\eqref{Eq:TMPC_sys} are satisfied by $\boldsymbol z_{[0,N]}$,  $\boldsymbol v_{[0,N-1]}$ and $a(k)=1$ for $P_k$. Notice that constraints \eqref{Eq:TMPC_SpecOld}-\eqref{Eq:TMPC_ProbGuar}, and \eqref{Eq:TMPC_SpecNew}-\eqref{Eq:TMPCNew_risk} in $P_{k-1}$ are equivalent to constraints \eqref{Eq:TMPC_SpecOld}-\eqref{Eq:TMPC_ProbGuar} in $P_k$ if $c(k)=1$. Hence, $\boldsymbol z_{[0,N]}$,  $\boldsymbol v_{[0,N-1]}$, $\boldsymbol \rho_{[0,N]}$ and $\boldsymbol r_{[0,N]}$ satisfy constraints \eqref{Eq:TMPC_SpecOld}-\eqref{Eq:TMPC_ProbGuar} for $P_k$. Finally, by taking $c(k)=1$ constraints \eqref{Eq:TMPC_SpecNew}-\eqref{Eq:TMPCNew_risk} are also  trivially satisfied. We conclude that $\{\boldsymbol z_{[0,N]}, \boldsymbol v_{[0,N-1]}, \boldsymbol \rho_{[0,N]}, \boldsymbol r_{[0,N]}, a(k), c(k)\}$ is a solution to $P_k$. This finishes the proof.
	\end{proof}
	
	\vspace{-0.9em}
	
	For each of the specifications $\psi\in\mathcal P^{N}$ in Algorithm \ref{Alg:ConRes}, the objective was to ensure that
	\begin{equation}
		\mathbb{P}_{\boldsymbol f}(\psi, k_\psi)\geq 1- r_{\psi,\max}.\label{eq:Psat_requirement}
	\end{equation}
	To show that this holds, it is sufficient to show that if at $k=k_{\psi}$, the optimal solution with $\boldsymbol r^{(k=k_{\psi})}_{[0,N]}$ is such that $$\textstyle \sum_{j \in \mathcal{H}_\psi^{k_\psi}\setminus \{k_\psi\}} \ r^{(k=k_{\psi})}(j) \leq  r_{\psi,\max},$$ then the open-loop implementation will satisfy \eqref{eq:Psat_requirement}. This would equate to solving \eqref{Eq:TMPC} and \eqref{Eq:TMPCNew} with $z(k)=z_k$. Additionally, as shown in \citet[Thm. 3]{Hewi2018}, if we instead use the updated state $z(k) = x(k)$, we still strictly preserve the original probability bounds due to the definition of the probabilistic reachable sets and the unimodal convexity of the additive disturbance.


	\section{Numerical Simulation} \label{Sec:App}

	Consider a two-dimensional robot-motion planning case to validate the efficacy of the proposed control rescheduling method. The dynamic model of the robot is described by Eq.~\eqref{Sys} with $A=B=I_2$, $x(0)=(0, 0)$, $n=2$, and $w(k) \sim \mathcal{N}(0,\sigma I_2)$ independent for all $k \in \{0, \cdots,N-1\}$, with $\sigma=0.002$. We consider a finite time-horizon $N=40$, a stabilizing feedback gain $K\approx -0.618 I_2$ for Eq.~\eqref{ErrorPart}, a maximal risk $r_{\psi, \max}=0.5$ for any specification $\psi$ and a cost $\textstyle J(\boldsymbol{z}, \boldsymbol{v}, \boldsymbol{r}) = \sum_{i=0}^{N-1}v(i)^TRv(i) + \sum_{i=1}^{N}r(i)$ with $R=0.001I_2$. A coordinate transformation is performed to ensure all PRS are spherical and dynamics normalized.
	
	Instead of the non-linear constraints \eqref{Eq:TMPC_ConRiskPRT}, we impose the following equality and inequality constraints, which imply that the original constraints hold. These constraints are computationally preferable, as they are quadratic convex equality and linear inequality constraints, respectively. We replace the non-linear equality \eqref{Eq:TMPC_ConRiskPRT} with a quadratic equality $r(k) = a \rho(k)^2 +b$ and a linear inequality $0.01 \leq r(k) \leq 1$, with $a=-0.005$ and $b=1.01$. It is easy to verify that $a \rho(k)^2 +b \geq \frac{n}{\rho(k)^2}$ always holds for $r(k) \in [\,0.01, 1\,]$. 
	
\begin{figure}[htp]
		\includegraphics[width=0.9\columnwidth]{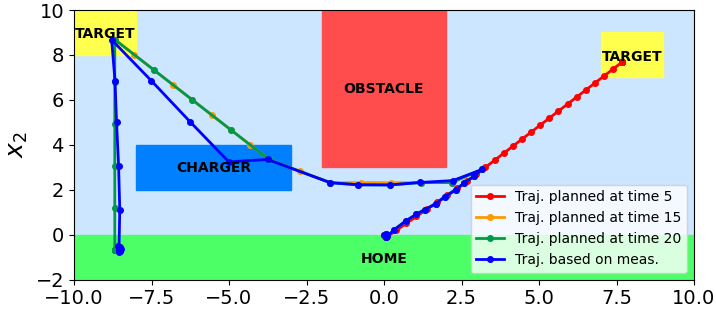}
		\centering
		\vspace{-0.8em}
		\caption{ The red, orange and green trajectory show robotic movement planned at $k=5$, $k=15$ and $k=20$ respectively. The blue trajectory shows the robots' actual movement.}
		\label{Fig:StateMeas}
	\end{figure}

	The motion planning scenario is illustrated in Fig.~\ref{Fig:StateMeas} which contains a safety set $S:=\{(x_1, x_2) \in \mathbb{R}^2 \mid -10 \leq x_1 \leq 10 \text{ and } -2 \leq x_2 \leq 10\}$, a home set $H:=\{(x_1, x_2) \in \mathbb{R}^2 \mid -10 \leq x_1 \leq 10 \text{ and } -2 \leq x_2 \leq 0\}$, an obstacle $O:=\{(x_1, x_2) \in \mathbb{R}^2 \mid -2 \leq x_1 \leq 2 \text{ and } 3 \leq x_2 \leq 10\}$, two targets $T_1:=\{(x_1, x_2) \in \mathbb{R}^2 \mid -10 \leq x_1 \leq -8 \text{ and } 8 \leq x_2 \leq 10\}$ and $T_2:=\{(x_1, x_2) \in \mathbb{R}^2 \mid 7 \leq x_1 \leq 9 \text{ and } 7 \leq x_2 \leq 9\}$, and a charger $C:=\{(x_1, x_2) \in \mathbb{R}^2 \mid -8 \leq x_1 \leq -3 \text{ and } 2 \leq x_2 \leq 4\}$. Note that all sets above are rectangular and can easily be represented as normalized polyhedral predicates.
	
	This case study considers a reach-avoid problem in which new reach objectives are dynamically assigned. More precisely, the specifications $\{\phi_0, \phi_1, \phi_2, \phi_3\}$ are assigned to the robot at times $k=0, \ k=5, \ k=15$ and $k=20$, respectively. Here $\phi_0:=\square_{[0,40]}\left[(x(k) \in S) \wedge \lnot (x(k) \in O)\right]$, $\phi_1:=\lozenge_{[20,30]}\left[(x(k) \in T_1) \vee (x(k) \in T_2)\right]$, $\phi_2:=\lozenge_{[20,25]}(x(k) \in C)$, and $\phi_3:=\lozenge_{[25, 30]}\square_{[0,5]}(x(k) \in H)$. Additionally, we specify that the control inputs are limited by $-2 \leq u(k) \leq 2$ for all $k \in \{0,1,\cdots, N-1\}$. We use the \textit{stlpy} toolbox to synthesize our controller~\citep{Kurt2022}. The results are shown in Fig. \ref{Fig:StateMeas} and Table \ref{tab:SpecLowBound}.

	\begin{table}[htp]
		\centering
		\resizebox{0.6\columnwidth}{!}{%
			\begin{tabular}{|l|c|c|c|c|}
				\hline
				& \multicolumn{1}{l|}{$\phi_0$} & \multicolumn{1}{l|}{$\phi_1$} & \multicolumn{1}{l|}{$\phi_2$} & \multicolumn{1}{l|}{$\phi_3$} \\ \hline
				$k=k_i$ & 0.40                           & 0.10                           & 0.05                          & 0.10                           \\ \hline
				$k=N$  & 0.40                           & 0.10                           & 0.05                          & 0.10                           \\ \hline
			\end{tabular}%
		}
		\vspace{0.5em}
		\caption{ Upper bound risks of $\phi_i$ at time of acceptance $k_i$ and after runtime.}
		\label{tab:SpecLowBound}
	\end{table}

	\vspace{-1em}
	
	From Fig. \ref{Fig:StateMeas}, our main observation is that at time $k=5$, the robot was scheduled to move towards the right target, only for it to move to the left target when it was additionally tasked with recharging its battery at time $k=15$. This result clearly illustrates that our control scheme allows the robot to make changes during runtime to accomplish as many tasks as possible. Additionally, open-loop guarantees on specifications, see Table \ref{tab:SpecLowBound}, remained below the maximum of $r_{\psi,\max}=0.5$. This result illustrates the ability to obtain valid guarantees.
	



	\section{Conclusion}\label{sec:conclu}

	In this paper, we developed a formal rescheduling control scheme for linear stochastic systems that must satisfy dynamically assigned specifications while preserving the performance of existing specifications. This was achieved by integrating real-time specification updates with probabilistic reachable tubes, risks and tube-MPC algorithms. A limitation of the developed approach is the conservativeness introduced in the upper bound risks due to the probabilistic reachable tubes. Relaxing this further will be considered in future work. We will also investigate an extension of the method to multi-agent systems, where specifications rejected by one agent are assigned to other agents within the system. 

	
	\if\mycmd0
	\section*{Appendix}

	\begin{lemma}[Risk-Aware Tube Lemma] \label{Lem:Tube}
		The risk that errors $\{e(0), \cdots, e(N)\}$ are not contained within a given probabilistic reachable tube $\{\mathcal{R}^{p_0}, \cdots, \mathcal{R}^{p_N}\}$ is at most
		\begin{equation}
			\bar r =\sum_{i = 1}^N (1-p_i).
		\end{equation}
	\end{lemma}

	\begin{proof}
		The probability that the errors are contained within the PRT is given by $p=1-\mathbb{P}(\cup_{i=0}^N E_i)$, where $E_i$ is the event that the $i^{\text{th}}$ element of the PRT does \emph{not} contain the $i^{\text{th}}$ element of the errors. Notice that $\mathbb{P}(E_0)=0$ and $\mathbb{P}(E_i)\leq r_i=1-p_i$. We can lower bound $p$ as follows:
		\begin{equation*}
			\bar p=1-\sum_{i =1}^N r_i\leq1-\sum_{i = 0}^N \mathbb{P}(E_i)\leq  1-\mathbb{P}(\cup_{i=0}^N E_i) =p.
		\end{equation*}
		\normalsize
		The first inequality is due to the probabilistic reachable set relation \eqref{Eq:PRT}, and the second is due to the union bound argument or Boole's inequality \citep{Bool1847}. An upper bound on the risk is then obtained from $\bar r=1-\bar p$.
	\end{proof}

	\noindent \textit{Explanation and Proof of Theorem \ref{Thm:Risk}. }

	\textbf{From robustness per time instance $i$ to risk. } Note that $\rho(i)$ performs a similar function as the robustness notion in \citet{Donz2010}. However, unlike those notions, we have opted for a local notion of robustness for the $i$-th time instance. That is, $\rho(i)$ defines the slack available in the constraints active for the state $x(i)$. Furthermore, given that the predicates are normalized, $\rho(i)$ denotes the minimal distance between the state $x(i)$ and the active constraints. This crucially determines the maximal radii of the probabilistic reachable tube for which the constraints are still satisfied. Thereby, we can quantify the risk $r_{\phi}^k$ as given in Thm. \ref{Thm:Risk}.

	\begin{proof}
		The nominal trajectory $\mathbf{z}_{[k,N]}$ satisfies the specification $\phi$ since $\rho(i)$ is positive for all $i$. We now need to prove the lower bound of the satisfaction probability based on the actual trajectory $\boldsymbol x_{[k,N]}$ that is affected by the noise. Given the sequence of $\rho$ values we know that $\boldsymbol x_{[k,N]}$ satis-fies the specification if $\forall i \in \mathcal{H}_{\phi}^k \backslash \{k\}$ the error $e(i)$ is con-strained as  $\|e(i)\|\leq \rho(i)$. Based on Lemma \ref{Lem:Tube} and Eq. \eqref{Eq:RelRadProb}, we have than that $\mathbb{P}_{\boldsymbol f}(\phi,k)\geq 1-r_{\phi}^k$ with $r_{\phi}^k$  as in \eqref{Eq:Risk}.
	\end{proof}

	\noindent \textit{Additional information on minimal distances. }
	
	The minimal distance between a point $\bar x$ and a normalized hyperplane $g^Tx=b$, i.e., $||g||=1$, is given by $l=|g^T\bar x-b|$.
	
	\noindent \textit{Additional information on the derivations of \eqref{Eq:STLCon}. }
	
	In \citet[IV-B]{Rama2014}, each predicate $\mu$ can be represented by two inequality constraints and a binary variable $s_k^\mu$. Here, the constraints enforce that $s_k^\mu=1$ if the predicate is satisfied at time step $k$, i.e. $h(\boldsymbol x(k))\geq \boldsymbol 0$. Utilizing the notation of this paper with minor simplification, these inequality constraints are given by
	\begin{subequations}
		\begin{align}
			h(\boldsymbol x(k)) &\geq (M(s_k^{\mu}-1)+\epsilon)\boldsymbol{1}_q,\\
			-h(\boldsymbol x(k)) &\geq (\epsilon-Ms^\mu_k)\boldsymbol{1}_q,
		\end{align}
	\end{subequations}
	where $M$ is a large positive number and $\epsilon$ is a small positive number. Notice that, should the predicate be satisfied at time $k$, $h(\boldsymbol x(k)) \geq \epsilon\boldsymbol{1}_q$. By replacing $\epsilon$ with $\rho(k)$, we can also consider the minimal distance between the state $\boldsymbol x(k)$ and the hyperplane boundaries of the normalized polyhedron predicate. To ensure the constraints remain valid, we assume that $\epsilon \leq \rho(k)\leq \frac{M}{2}$, ensuring that $\rho(k)-M$ is a large negative number, similar to $\epsilon-M$.
	\fi	
	
	\bibliography{Article}

\end{document}